\documentclass[10pt, twocolumn, twoside, a4paper]{article}

\usepackage[left=1.5cm, right=1.5cm, top=2cm, bottom=2cm, columnsep=0.6cm]{geometry}

\usepackage{cite}
\usepackage{amsmath, amsbsy, amsfonts, amssymb, amscd, amsthm}
\usepackage{algorithmic}
\usepackage{graphicx}
\usepackage{textcomp}
\usepackage{xcolor}
\usepackage{authblk}

\usepackage{fancyhdr}
\newtheorem{theorem}{Theorem}[section]

\newtheorem{remark}[theorem]{Remark}

\newcommand{\be}{\begin{equation}}
\newcommand{\ee}{\end{equation}}
\newcommand{\bea}{\begin{eqnarray}}
\newcommand{\eea}{\end{eqnarray}}

\newcommand{\bml}{\begin{mathletters}}
\newcommand{\eml}{\end{mathletters}}

\def\BibTeX{{\rm B\kern-.05em{\sc i\kern-.025em b}\kern-.08em
    T\kern-.1667em\lower.7ex\hbox{E}\kern-.125emX}}

\begin{document}

\title{A Generalized Framework of Antisymmetric Polyspectral Indices for Identifying High-Order Neural Interactions}
\author[1,2]{Alessio Basti}
\author[3]{Rikkert Hindriks}
\author[4]{Ruggero Freddi}
\author[2]{Gian Luca Romani}
\author[2,5]{Vittorio Pizzella}
\author[6,*]{Guido Nolte}
\author[1,2,*]{Laura Marzetti}

\affil[1]{Department of Engineering and Geology, ``G. d'Annunzio" University Chieti-Pescara, Italy}
\affil[2]{Institute for Advanced Biomedical Technologies, ``G. d'Annunzio" University Chieti-Pescara, Italy}
\affil[3]{Department of Mathematics, Vrije Universiteit Amsterdam, The Netherlands}
\affil[4]{Research and Development Department, Manava.plus, Italy}
\affil[5]{Department of Neuroscience, Imaging and Clinical Sciences, ``G. d'Annunzio" University Chieti-Pescara, Italy}
\affil[6]{Department of Neurophysiology and Pathophysiology, University Medical Center Hamburg-Eppendorf, Germany}
\affil[*]{These authors contributed equally to this work.}

\maketitle

\begin{abstract}
Cross-frequency interactions are fundamental brain mechanisms for integrating information across temporal scales. However, accurate identification of these couplings is hindered by complex multi-frequency nonlinearities and by spurious, zero-lag artifacts caused by volume conduction. To our knowledge, conventional metrics lack a robust framework to characterize genuine interactions among multiple time series where a frequency of interest $f_N$ arises from the combination of $N-1$ components such that $f_N = \sum_{i=1}^{N-1} f_i$. We introduce a general family of antisymmetric cross-polyspectral indices designed to quantify these harmonic dependencies while being intrinsically robust to instantaneous mixing. We derive the theoretical properties of these quantities and validate them through simulations of cubic nonlinearities. As a proof of concept, we apply the indices to empirical EEG recordings; the results reveal significant higher-order dependencies that elude standard analytical approaches. We further discuss how these indices can inform novel, personalized multi-site transcranial magnetic stimulation (mTMS) protocols by enabling the selective monitoring and modulation of specific multi-frequency network interactions.
\end{abstract}

\section{Introduction}
The brain can be conceptualized as a complex dynamical system whose macroscopic behavior emerges from the coordinated interplay of mesoscopic activities. At this intermediate scale, spatially distributed cortical and subcortical areas are indeed anatomically connected by long‑range projections and form functional networks that dynamically reconfigure to support perception, cognition, and action \cite{Bassett2017,Venkadesh2021}. Importantly, these inter‑areal interactions selectively recruit specific oscillatory components rather than the full broadband activity of each region, yielding an ensemble of frequency‑tagged processes intricately coupled both within (intra‑area) and between (inter‑area) regions \cite{Marzetti2013, Ji2019, Canolty2010}.

To track these rapid functional reorganizations with millisecond precision, non-invasive electrophysiology, e.g. electroencephalography (EEG) and magnetoencephalography (MEG), provides the necessary temporal resolution \cite{Bressler2010, Favaretto2022}. However, the utility of these techniques is often constrained by the metrics used to analyze the associated time series. Most conventional connectivity measures, such as the Phase-Locking-Value (PLV), the Imaginary part of the Coherency (ImCoh), and the weighted Phase Lag Index (wPLI), assume same-frequency interactions, effectively treating frequency bands in isolation \cite{Marzetti2019}.

 This "single-frequency" approach contrasts with growing evidence that neural computation relies on cross-frequency coupling (CFC) to coordinate processes across multiple temporal scales \cite{Canolty2010, Buzsaki2004}. Neglecting these dependencies risks missing the complex dynamical motifs that support information transfer and, crucially, those targeted by modern neuromodulation.

The need for robust CFC metrics is particularly pressing with the advent of multi-locus TMS (mTMS; \cite{Rissanen2023}). Unlike conventional systems, mTMS utilizes overlapping coil arrays to enable rapid, multi-site E-field manipulation without mechanical movement \cite{Koponen2018, Nieminen2022, Sinisalo2024}. By theoretically engaging multiple cortical loci at distinct rates $(f_1,…,f_{N-1})$, these paradigms can induce emergent drives at composite frequencies in areas of overlap:

\begin{equation}\label{tms} f_{\text{overlap}} = \sum_{i=1}^{N-1} f_i \end{equation}	

\begin{figure}
\centering
\includegraphics[width=0.48\textwidth]{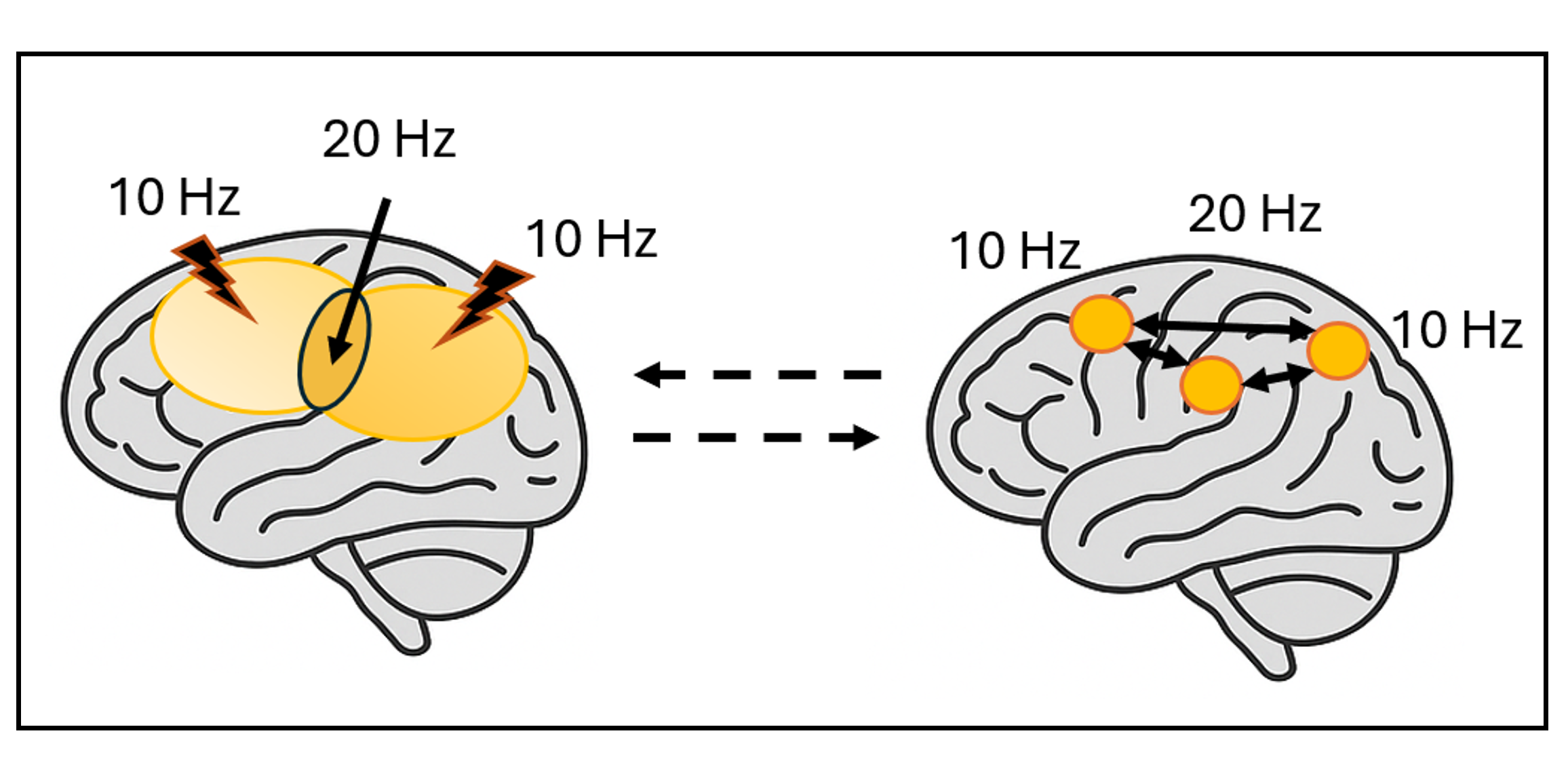}
\caption{Stimulating \(N\) partially overlapping cortical loci via multi-locus TMS at rates
\(f_1,\dots,f_{N-1}\) can induce composite spectral drive e.g. in the region of full overlap
(\(f_{\text{overlap}}\!=\!f_1+\dots+f_{N-1}\)).
Restricting stimulation control and feedback to same-frequency connectivity may miss
crucial dynamical features that are either altered by stimulation or actively drive
network reconfiguration. Estimating cross-frequency couplings therefore provides a
principled basis for monitoring these effects and for guiding personalized
multi-site / multi-frequency stimulation protocols.}
\label{fig:statcheck}
\end{figure}
 
In such scenarios, monitoring only same-frequency connectivity may fail to capture the emergent cross-frequency phenomena that actually mediate the stimulation's effects. Explicitly estimating such dependencies is therefore essential to inform and guide personalized, multi-site stimulation strategies.

Methodologically, identifying these interactions in EEG/MEG is hindered by volume conduction and field spread, which introduce spurious couplings through linear instantaneous mixing. While third-order spectral tools like the bispectrum can detect quadratic (1:2) interactions, they must be adapted for robustness. The Antisymmetric Cross-Bicoherence (ACB) was developed to suppress these mixing artifacts by exploiting the symmetry properties of the estimators \cite{Chella2014}. Recently, this was extended to multivariate signals (MACB) to improve statistical efficiency in high-dimensional source-level analyses \cite{Basti2024}.

Despite these advances, a robust framework for identifying genuine CFC beyond quadratic order, such as 1:3 (e.g., alpha-to-low-gamma) or among $n$ time series with frequencies such that Eq. (\ref{tms}) holds, is still lacking. Existing $n$:$m$ phase-locking or phase-amplitude coupling (PAC) measures often lack intrinsic robustness to linear mixing and tend to separate phase and amplitude, even though weighting phase by amplitude could improve the signal-to-noise ratio (SNR). Furthermore, these measures are typically defined only for region-pairs, do not naturally generalize to interactions among more than two regions, and lack a direct extension to multi-dimensional frameworks (e.g., parcel signals, 3$D$ source signals) \cite{Siebenhühner2020}.

In this article, we introduce a general family of antisymmetric cross-polyspectral indices designed to quantify 1:$n$ harmonic interactions among time series while remaining robust to field spread. We first derive a fourth-order statistic for 1:3 interactions, then generalize it to an $n$-th order formulation. We validate index performance with a generative model that exhibits nonlinear (e.g. cubic) interactions and systematically varies signal-to-noise ratio to probe it under realistic noise conditions. When applied to empirical single-subject EEG, the proposed indices reveal statistically significant higher-order cross-frequency dependencies that are observable directly at sensor level, underscoring the method’s practical applicability.

\section{Theoretical formulation}
In this section, we develop the framework underpinning our family of antisymmetric cross-polyspectral indices. We begin by recalling the classical definitions and then introduce an antisymmetrization procedure that formally cancels all contributions from independent linear sources, thereby isolating genuine coupling. For the sake of clarity and to avoid the excessive notational complexity of the general case, we first derive the fourth-order normalized measure, the Antisymmetric Cross-Tricoherence (ACT). We specifically focus our derivation on the bivariate case (two signals), and using $f_1=\dots=f_{N-1}=f$; however, this construction extends seamlessly to the multivariate scenario involving $n$ time series governed by Eq. (\ref{tms}). Building on this foundation, we finally show how the framework generalizes to interactions of arbitrary order.

\subsection{Antisymmetric cross-trispectrum}

We aim to define a trispectral index, \(T_{[x|xx|y]}\), for EEG/MEG time series \(x(t)\) and \(y(t)\). This index will detect cubic dependencies between the two signals while remaining robust to linear instantaneous mixing artifacts. Suppose that
\(
x(t) = \sum_{i=1}^M a_i s_i(t)
\)
and that
\(
y(t) = \sum_{i=1}^M b_i s_i(t),
\)
where \(s_i(t)\) are the zero-mean real-valued source signals, \(\{a_i\}_{i=1,...,M}\) and \(\{b_i\}_{i=1,...,M}\) are mixing coefficients. Henceforth, we omit explicit reference to the number of sources $M$, which is assumed to be finite throughout the article.
\begin{remark}
If \(x\) and \(y\) correspond to EEG/MEG sensor/channel signals, the linear mixing model accounts for volume‑conduction and field‑spread effects. Instead, if they represent source‑reconstructed activities, the mixing models the fact that, even after the application of an electromagnetic pseudo‑inverse method, a residual contribution persists due to the inherent limits of the procedure.
\end{remark}

We define the trispectrum as
\begin{equation}
\begin{aligned}
T_{xxxy} \;:&=\; \bigl\langle X(f)^3\,Y(3f)^* \bigr\rangle\\
&=\sum_{i,j,k,l}a_i a_j a_k b_l \,\bigl\langle S_i(f)\,S_j(f)\,S_k(f)\,S_l(3f)^*\bigr\rangle
,
\end{aligned}\end{equation}
where capital letters denote the Fourier transforms, and \(f\) is a frequency of interest. For the derivation below, we assume that the source processes $s_i(t)$ are also mutually independent, and strictly stationary, and that the expectations are taken in the population limit. Under these assumptions, mixed-source moments factorize, and the Fourier coefficients obey the standard spectral selection rules: a non-conjugated Fourier moment is nonzero only when the signed frequencies balance, and a conjugated moment is nonzero only when the positive and negative frequencies match.

Expanding the expectation under source independence factorizes each term into products of moments associated with individual sources. For the frequency configurations considered here, any mixed-source contribution contains at least one factor that violates the spectral-balance condition for a real-valued strictly stationary process, and therefore vanishes. For example, terms of the form
\(
\bigl\langle S_i(f)^3\,S_l(3f)^*\bigr\rangle
\)
factorize into
\(
\langle S_i(f)^3\rangle \,\langle S_l(3f)^*\rangle = 0,
\)
since each source has zero mean. Likewise, for paired terms one obtains factors such as
\(
\langle S_i(f)^2\rangle\) or \(
\langle S_l(f)\,S_l(3f)^*\rangle,
\)
and these vanish at nonzero frequencies under the spectral-selection rule.

Hence, under the null model of independent sources linearly mixed at the sensors, only same-source contributions remain. In particular,
\begin{equation}
\begin{aligned}
T_{xxxy}(f)
&=
\sum_i a_i^3 b_i \,\bigl\langle S_i(f)^3 S_i(3f)^*\bigr\rangle.
\end{aligned}
\end{equation}

The latter summation does not vanish without additional assumptions. We therefore define an index that, in the case of independent linear mixing, matches Eq. (3) exactly, and call the difference between these expressions the \textit{antisymmetric cross‑trispectrum}. Specifically, we set this as $T_{[x|xx|y]}=T_{xxxy}-T_{yxxx}$, where the second term is obtained by swapping the first and last indices. Indeed, 

\begin{equation}
\begin{aligned}
&T_{yxxx}=\sum_i a_i^3 b_i \,\bigl\langle S_i(f)^3\,S_i(3f)^*\bigr\rangle+\\
  &+\sum_{\substack{i,l\\i\neq l}} a_i^2 a_l b_l[
    \langle S_l(f)^2\,S_i(f)\,S_i(3f)^*\rangle+\\
  &+ \langle S_l(f)\,S_i(f)\,S_l(f)\,S_i(3f)^*\rangle + \langle S_l(f)\,S_i(f)^2\,S_l(3f)^*\rangle]\\
=& \sum_i a_i^3 b_i \,\bigl\langle S_i(f)^3S_i(3f)^*\bigr\rangle
  \end{aligned}
\end{equation}
  
Hence, under the null model, the antisymmetric contrast cancels the symmetric mixing contribution. 
\begin{remark}
In the formulation of the trispectrum we restrict ourselves, for notational simplicity, to the case of three identical frequencies \(f\) with the fourth frequency fixed to \(3f\). This restriction does not limit the framework: the construction generalizes to arbitrary triplets \(f_1,f_2,f_3\) and their sum \(f_1+f_2+f_3\), provided degenerate configurations are avoided. In particular, choosing \(f_2=-f_1\) yields
\[
f_1+f_2+f_3 = f_3,
\]
and produces non-vanishing contributions (even after antisymmetrization) of the form
\[
\big\langle S_i(f_1)\,S_i(-f_1)\,S_j(f_3)\,S_j(f_3)^*\big\rangle
\]\[=
\big\langle S_i(f_1)\,S_i(f_1)^*\big\rangle
\;\big\langle S_j(f_3)\,S_j(f_3)^*\big\rangle,
\]
as discussed in Chella et al.\ (2014). Such factorable terms undermine robustness to linear mixing; however, this degeneracy is essentially irrelevant in practice because neural signal processing typically consider only positive frequencies. Therefore, excluding these configurations (which is standard practice) makes the extension to arbitrary frequency triplets straightforward.

\end{remark}
\subsection{Univariate normalization}

In order to bound the magnitude of the antisymmetric cross‑trispectrum within the interval \([0,1]\), we introduce a normalization following \cite{Shahbazi2014} and \cite{Chella2016}.  

First, define for each time series the fourth‑order amplitude norm
\begin{equation}
Q_x(\tilde f)
=\Biggl(\big\langle\bigl|X(\tilde f)\bigr|^4\big\rangle\Biggr)^{1/4},
\end{equation}
and similarly \(Q_y(\tilde f)\). In Eq. (5), $\tilde f$ denotes the generic frequency. Then each term of \(T_{[x|xx|y]}\) satisfies (e.g. via the Hölder bound)
\[
\bigl|T_{xxxy}(f)\bigr|\le Q_x(f)^3\,Q_y(3f),\]\[
\bigl|T_{yxxx}(f)\bigr|\le Q_y(f)\,Q_x(f)^2\,Q_x(3f).
\]

We therefore introduce the \emph{antisymmetric cross‑tricoherence} (ACT)
\[
\Gamma_{[x|xx|y]}^{(4)}(f)
\;=\;
\frac{T_{[x|xx|y]}}
{\,Q_x(f)^3\,Q_y(3f)\;+\;Q_y(f)\,Q_x(f)^2\,Q_x(3f)\,}\,.
\]
Indeed, by the triangle inequality and the individual Hölder bounds,
\[
\bigl|T_{[x|xx|y]}\bigr|
\le \bigl|T_{xxxy}\bigr| + \bigl|T_{yxxx}\bigr|
\le Q_x^3\,Q_y(3f) + Q_y\,Q_x^2\,Q_x(3f),
\]
hence
\[
\biggl|\Gamma_{[x|xx|y]}^{(4)}(f)\biggr|\le1.
\]
\begin{remark}
The superscript \((4)\) in $\Gamma_{[x|xx|y]}^{(4)}(f)$ indicates that this quantity is associated with a fourth‑order statistic, despite the term “trispectrum” suggesting third order. In reality, the trispectrum is related to the fourth‑order cumulant function. We adopt this notation both to emphasize the statistical order and to facilitate the extension to a similar index for a generic \(p\)-th order statistic. Furthermore, it is worth noting that from the very first Eq (1), we focus on a specific trispectrum \(T_{xxxy}\), involving three instances of the time series \(x\) and one of \(y\). Computationally speaking, for any four (not necessarily identical) time series \(\{x,y,w,z\}\), it is possible to define the associated trispectrum, generating a four-dimensional array.
\end{remark}

\subsection{Formal toy-example for $\bigl|\Gamma_{[x|xx|y]}^{(4)}(f)\bigr|=1$}
Now suppose that the $n$-th segment of a time series $x$ corresponds to a frequency-specific cosine with independent random phase \(\phi_n\), and that the one of $y$ is equal to its cubic version, i.e.
\[\begin{aligned}
x(t;n) &= A\cos\bigl(2\pi f t + \phi_n\bigr),\\
y(t;n) &= k\bigl[A\cos(2\pi f t + \phi_n)\bigr]^3\\
&= k\,A^3\,\frac{3\cos(2\pi f t + \phi_n) + \cos(2\pi\,3f\,t + 3\phi_n)}{4}.\end{aligned}
\]
Ideally, in the Fourier representation for each segment \(n\), one obtains
\[
X(f;n) = \frac{A}{2}\,e^{\,i\phi_n},
\quad
X(3f;n) = 0,\]\[
Y(f;n) = \frac{3k\,A^3}{8}\,e^{\,i\phi_n},
\quad
Y(3f;n) = \frac{k\,A^3}{8}\,e^{\,i3\phi_n},
\]
For the sake of simplicity, let us now avoid noting the segment with $n$. The normalization factors are equal to
   \[
   Q_x(f)
   = \Bigl(\tfrac1N\sum_{n}\Bigl|\tfrac{A}{2}\Bigr|^4\Bigr)^{1/4}
   = \frac{|A|}{2}, \quad Q_x(3f)=0\]\[
   Q_y(f)=\tfrac{3|k|\,|A|^3}{8}, \quad Q_y(3f)
   = \frac{|k|\,|A|^3}{8}.
   \]
Since each segment contributes a phase factor that exactly cancels in the product,
   \[\begin{aligned}
   T_{xxxy}(f)
   &= \frac{1}{N}\sum_{n}\Bigl(\tfrac{A}{2}e^{i\phi_n}\Bigr)^3
                     \Bigl(\tfrac{k\,A^3}{8}e^{i3\phi_n}\Bigr)^*\\
   &= \frac{k\,A^6}{2^6}\;\frac{1}{N}\sum_{n}e^{i(3\phi_n-3\phi_n)}
   = \frac{k\,A^6}{2^6}.
   \end{aligned}\]
   Likewise,
   \[
   T_{yxxx}(f)
   = \frac{1}{N}\sum_{n}\Bigl(\tfrac{3k\,A^3}{4}e^{i\phi_n}\Bigr)
                         \Bigl(\tfrac{A}{2}e^{i\phi_n}\Bigr)^2
                         0^*
   = 0.
   \]

Therefore 
   \[\begin{aligned}
   \biggl|\Gamma_{[x|xx|y]}^{(4)}(f)\biggr|
   &= \frac{|T_{xxxy}-T_{yxxx}|}
          {\,Q_x(f)^3\,Q_y(3f)+Q_y(f)\,Q_x(f)^2\,Q_x(3f)\,}\\
   &= \frac{\tfrac{|k|\,A^6}{2^6}-0}{\tfrac{|k|\,A^6}{2^6}}
   = 1.
   \end{aligned}\]
\subsection{Generalization to Antisymmetric Cross-Polyspectrum}

We now show that exactly the same construction extends to arbitrary order \(m\in \mathbb{N}\), \(m\geq2\).  Let \(x\) and \(y\) be two time series mixed from independent, zero-mean, real-valued, strict-sense stationary sources \(s_i(t)\).  Define the $m$-th order \emph{cross-polyspectrum} as
\[
P^{(m)}_{\,\underbrace{x\cdots x}_{m-1}\,y}(f)
:= \bigl\langle 
  \underbrace{X(f)\,\cdots\,X(f)}_{m-1\ \text{times}}\,
  Y\bigl((m-1)f\bigr)^*
\bigr\rangle\]
\[= \sum_{i_1,\dots,i_{m}}
  a_{i_1}\cdots a_{i_{m-1}}\,b_{i_{m}}
  \bigl\langle
    S_{i_1}(f)\cdots S_{i_{m-1}}(f)\,S_{i_{m}}((m-1)f)^*
  \bigr\rangle.
\]
By the same index-vanishing arguments as before, only terms with \(i_1=\cdots=i_{p}\) survive, so
\[
P^{(m)}_{\,x\cdots x\,y}(f)
= \sum_i a_i^{m-1}\,b_i\;\bigl\langle S_i(f)^{m-1}\,S_i((m-1)f)^*\bigr\rangle.
\]
To cancel any symmetric mixing contributions, we compare
\[
P^{(m)}_{\,x\cdots x\,y}
\quad\longleftrightarrow\quad
P^{(m)}_{\,y\,x\cdots x}
\]
where the latter is obtained by permuting the first and last factors. By the same combinatorial proof, one shows
\[
P^{(m)}_{\,y\,x\cdots x}(f)
= \sum_i a_i^{m-1}\,b_i\;\bigl\langle S_i(f)^{m-1}\,S_i((m-1)f)^*\bigr\rangle.
\]
Hence, the $m$-th order \emph{antisymmetric cross-polyspectrum}
\(
P^{(m)}_{[\,x|x\cdots x\,|\,y\,]}
:= P^{(m)}_{\,x\cdots x\,y} \;-\; P^{(m)}_{\,y\,x\cdots x}\,
\) vanishes under the hypotheses described above. By introducing the \((m)\)-th order amplitude norm
\[
Q_x(f)
:= \Bigl\langle \bigl|X(f)\bigr|^{m}\Bigr\rangle^{1/m},
\]
we have \[\begin{aligned}
\bigl|P^{(m)}_{\,x\cdots x\,y}(f)\bigr|
&\le \underbrace{Q_x(f)\,\cdots\,Q_x(f)}_{m-1\ \text{times}}\,Q_y((m-1)f)\\
&= Q_x(f)^{m-1}\,Q_y((m-1)f),
\end{aligned}\]
and likewise
\(\bigl|P^{(m)}_{\,y\,x\cdots x}(f)\bigr|\le Q_y(f)\,Q_x(f)^{m-2}\,Q_x((m-1)f)\).  Therefore we define the $m$-th order \emph{antisymmetric cross-polycoherency} (ACP)
\[\begin{aligned}
&\Gamma^{(m)}_{[\,x|x\cdots x\,|\,y\,]}(f)
:=\\ 
&\frac{P^{(m)}_{[\,x|x\cdots x\,|\,y\,]}(f)}
        {\,Q_x(f)^{m-1}\,Q_y((m-1)f)\;+\;Q_y(f)\,Q_x(f)^{m-2}\,Q_x((m-1)f)\,}\,,
\end{aligned}\]
and by the triangle inequality,
\(
\bigl|\Gamma^{(m)}_{[\,x|x\cdots x\,|\,y\,]}(f)\bigr|
\le 1.
\)
\subsection{Special case \(m=2\): the second order ACP is the imaginary coherency}

Setting \(m=2\) in the general definitions, we have
\[
P^{(2)}_{xy}(f)\;=\;\bigl\langle X(f)\,Y(f)^*\bigr\rangle,
\]\[
P^{(2)}_{yx}(f)\;=\;\bigl\langle Y(f)\,X(f)^*\bigr\rangle
=P^{(2)}_{xy}(f)^*.
\]
Hence, by swapping the indices,
\[\begin{aligned}
P^{(2)}_{[x|y]}(f)
&= P^{(2)}_{xy}(f)\;-\;P^{(2)}_{yx}(f)\\
&= \langle X\,Y^*\rangle \;-\;(\langle X\,Y^*\rangle)^*
= 2\,i\,\Im\bigl\langle X(f)\,Y(f)^*\bigr\rangle.
\end{aligned}\]

The normalization factors for \(m=2\) are
\[
Q_x(f)
= \bigl\langle |X(f)|^2\bigr\rangle^{1/2},
\qquad
Q_y(f)
= \bigl\langle |Y(f)|^2\bigr\rangle^{1/2},
\]
and the denominator becomes
\[
Q_x(f)\,Q_y(f)\;+\;Q_y(f)\,Q_x(f)^0\,Q_x(f)
= 2\,Q_x(f)\,Q_y(f).
\]
Therefore,
\[
\Gamma^{(2)}_{[x|y]}(f)
= \frac{P^{(2)}_{[x|y]}(f)}{2\,Q_x(f)\,Q_y(f)}
= i\,\frac{\Im\langle X(f)\,Y(f)^*\rangle}{Q_x(f)\,Q_y(f)}.
\]
Taking the magnitude yields
\[
\biggl|\Gamma^{(2)}_{[x|y]}(f)\biggr|
= \frac{\bigl|i\,\Im\langle X(f)\,Y(f)^*\rangle\bigr|}{Q_x(f)\,Q_y(f)}
= \frac{\bigl|\Im\langle X(f)\,Y(f)^*\rangle\bigr|}{\sqrt{\langle|X(f)|^2\rangle}\,\sqrt{\langle|Y(f)|^2\rangle}},
\]
which is the absolute value of the \emph{imaginary part of the coherency} between \(x\) and \(y\) \cite{Nolte2004}. Similarly, the antisymmetric cross-bicoherence \cite{Chella2016} corresponds to the case $m=3$.

\section{Experiments}
\subsection{Synthetic Data Experiment}
\label{sec:simulations}
The theoretical properties of the estimators were assessed through simulations across a range of signal-to-noise ratios, focusing on cubic nonlinearity regimes (1:3 cross-frequency coupling) via the proposed antisymmetric cross-tricoherence (ACT). The simulations are designed to reproduce two neural/neuroengineering phenomena, which may be encountered in EEG/MEG: (i) a genuine cubic coupling in which one time series carries a cubic nonlinearity of a base oscillation present in another time series, and (ii) artifactual instantaneous mixing produced by correlated noise sources linearly combined at the sensors. By interpolating between these two extremes via a scalar parameter $\gamma$ we sweep effective signal-to-noise conditions and quantify the robustness of ACT as compared to non-antisymmetrized trispectral measures. Our MATLAB implementation followed the workflow below.

\subsubsection{Generative model}
Let $t$ index discrete time samples. Two observed time series (channels) $x(t)$ and $y(t)$ are built as convex combinations of a “coupled” component and a “linear mixture” (artifact) component:
\begin{equation}\label{eq:mixing_interp}
\begin{aligned}
x(t;\gamma) &= (1-\gamma)\,\tilde{s}_x(t) + \gamma\,\tilde{\ell}_x(t),\\
y(t;\gamma) &= (1-\gamma)\,\tilde{s}_y(t) + \gamma\,\tilde{\ell}_y(t),
\end{aligned}
\end{equation}
where $\gamma\in[0,1]$ controls the correlated-noise weight. Signals with the tilde are \emph{normalized} (see below); $\tilde{s}_x,\tilde{s}_y$ are related to the genuine coupled sources, and $\tilde{\ell}_x,\tilde{\ell}_y$ denote linear mixtures of independent noise sources that simulate artifactual instantaneous interactions.

The coupled components are obtained from a single band-limited oscillator $u(t)$:
\begin{equation}\label{eq:coupled}
\begin{aligned}
u(t) &= \text{bandpass}\bigl(\eta(t); f_{\mathrm{low}}, f_{\mathrm{high}}\bigr),\quad \eta(t)\sim\mathcal{N}(0,1),\\
s_x(t) &= u(t-\tau),\\
s_y(t) &= \bigl(u(t)\bigr)^3,
\end{aligned}
\end{equation}
where $\tau$ is an integer lag (in samples) that creates a delayed relationship between $x$ and $y$. In the MATLAB experiments we set $f_{\mathrm{low}}=5\,$Hz, $f_{\mathrm{high}}=15\,$Hz and $\tau$ equal to 10 samples (with a simulated sampling rate of 256 Hz). The cubic nonlinearity produces power at the base frequency $f$ and at its third harmonic $3f$, implementing an idealized $1\!:\!3$ cross-frequency coupling.

The artifactual linear mixtures are generated from three independent band-limited noise sources $n_k(t)$ (for $k=1,2,3$), each cubed to mimic higher-order spectral content:
\begin{equation}\label{eq:noise_lin}
n_k(t)=\bigl[\text{bandpass}(\xi_k(t); f_{\mathrm{low}}, f_{\mathrm{high}})\bigr]^3,\qquad \xi_k(t)\sim\mathcal{N}(0,1),
\end{equation}
and the linear mixtures are
\begin{equation}
\ell_x(t) = \sum_{k=1}^3 a_k\, n_k(t),\qquad
\ell_y(t) = \sum_{k=1}^3 b_k\, n_k(t),
\end{equation}
with random coefficients $a_k,b_k\overset{\text{iid}}{\sim}\mathcal N(0,1)$. The cube operation on the noise sources was included to populate higher-order spectral components while preserving the linear instantaneous mixing structure.

Finally, before convex interpolation (Eq.~\ref{eq:mixing_interp}) each component vector is normalized (\(2\)-norm), i.e.
\[
\tilde{s}_x = \frac{s_x}{\|s_x\|_2},\quad
\tilde{\ell}_x = \frac{\ell_x}{\|\ell_x\|_2},
\]
(and analogously for $\tilde{s}_y,\tilde{\ell}_y$). This ensures that $\gamma$ directly parameterizes the relative contribution of the artifact versus the genuine coupling independent of scale.

\subsubsection{Trispectral estimation and summary statistics}
Trispectral quantities are estimated in the usual block-averaged manner. Given a data segment of length $L$ samples, we apply a window $w[n]$ (Hanning), compute discrete Fourier transforms on overlapping segments, and then use the formulas as above.  

For comparison, in addition to the antisymmetric cross-tricoherence (ACT) estimator, $\Gamma_{[x|xx|y]}^{(4)}$, we also compute the normalized versions of $T_{xxxy}$ and $T_{yxxx}$ (referred to as CT1 and CT2 indices, respectively).

To aggregate results across the ($N_{iter}=1000$) independent runs, and to reduce sensitivity to outlying runs, we \emph{normalize each run individually} by dividing the measured curve (ACT or CT versus $\gamma$) by its run-wise maximum, so that each run provides values in $[0,1]$. Let $\Gamma^{(r)}(\gamma)$ denote the normalized value at $\gamma$ for run $r$. Across runs we then compute the pointwise median $\mathrm{median}_r\{\Gamma^{(r)}(\gamma)\}$ as a robust central tendency estimator, and the pointwise interquartile range (IQR), given by the 25\textsuperscript{th} and 75\textsuperscript{th} percentiles, to quantify variability across runs.

For visualization we display the median curve together with a shaded band spanning [25\%,75\%] (IQR). To allow direct comparison across measures, the vertical axis is bounded above by $1$ and set below to a common minimum $y_{\min}$ determined as the smallest 25\% percentile observed across the considered measures.

\begin{figure}
\centering
\includegraphics[width=0.48\textwidth]{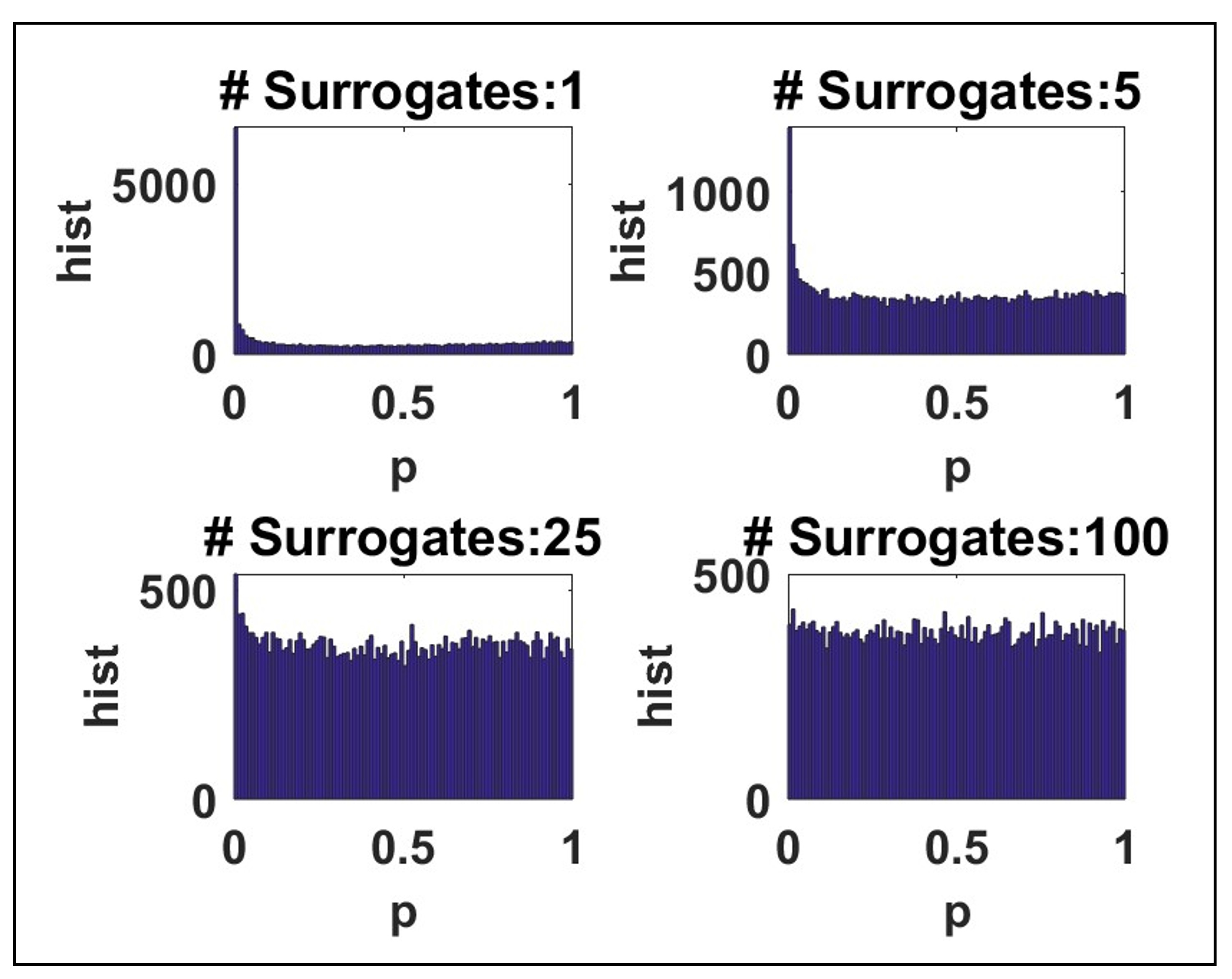}
\caption{Histograms of $p$-values for surrogate tests. The distributions shown clearly approach uniformity as $N$ (number of surrogate datasets for each original dataset) increases, with the case $N=100$ closely matching the expected uniform profile.}
\label{fig:statcheck}
\end{figure}
\subsection{Real EEG Data Experiment}

As a showcase we present results for an empirical single subject dataset freely available from the METH toolbox, consisting of about 6 minutes of resting state EEG data with 61 channels \cite{Goschl2025}. 

\subsubsection{A statistics for trispectra}

We study 3 different variants of a trispectrum: CT1, CT2 and ACT. For the statistical evaluation of trispectral estimates we follow the approach of \cite{Bartz2020}. This is illustrated for one specific measure but applies equally to any other of those measures. Let $x_{ik}(f)$ be the Fourier coefficient of the signal in sensor $i$ at frequency $f$ for the $k.th$ segment, and the measure discussed here is

\be
z=\frac{1}{K}\sum_{k=1}^K x_{ik}(f_1)x_{ik}(f_2)x_{ik}(f_3)x_{jk}^*(f_1+f_2+f_3)
\ee
for $K$ segments. 
The question is whether the empirical value $|z|$ is inconsistent with the null-hypothesis that there is no cross-frequency coupling observable with this measure. To assess this we can generate $N$ surrogate data sets by randomly permuting the temporal order of the segments  of the fourth Fourier coefficient. Specifically, let $\Pi_n$ be the $n$-th permutation for $n=1, \dots, N$ , then the $n$-th measure for the surrogates reads
\be
\hat{z}(n)=\frac{1}{K}\sum_{k=1}^K x_{ik}(f_1)x_{ik}(f_2)x_{ik}(f_3)x_{j\Pi_n(k)}^*(f_1+f_2+f_3)
\ee 
If there is no cross-frequency coupling, the relative temporal order of the segments is statistically irrelevant, and $z$ follows the same distribution as $\hat{z}(n)$.  

For the parametric test we exploit the fact that, being an average over a large number of segments, and assuming a uniform phase distribution for each individual Fourier coefficient, $\hat{z}(n)$ is circular symmetric and approximately Gaussian distributed with zero mean. Then the absolute value is Raleigh distributed. We first calculate  
the quantity 
\be
r=\frac{|z|^2}{\frac{1}{N}\sum_n|\hat{z}(n)|^2}
\label{eq:r}
\ee    
and  the p-value can be approximated as 
\be
p=\exp(-r).   
\ee

As a sanity check for the parametric statistics we replaced the empirical data with random white noise data of the same size and analyzed 10 such datasets for all channel pairs. For 61 channels we get a total of $10*61^2=37210$ p-values. For each original dataset we constructed $N$ surrogate datasets for $N=1,5,25,100$. If the statistics is correct, the estimated p-values should be uniformely distributed. The observed distribution is shown in Figure \ref{fig:statcheck}.

We observe that for few surrogate data sets small $p$-values occur too often. This effect becomes smaller for a larger number of surrogate data sets but it never vanishes completely. 
\begin{figure*}[!t]      
  \centering
  \includegraphics[width=480pt]{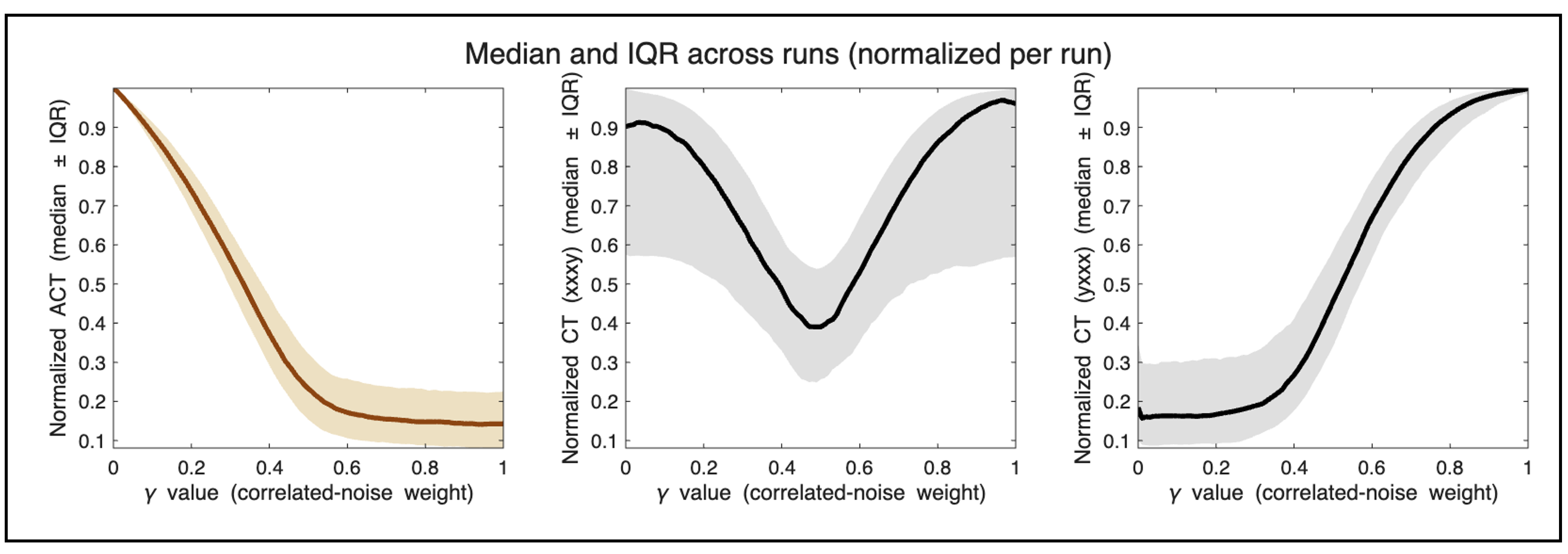} 
  \caption{Left: ACT (median ± IQR) remains sensitive to genuine cubic coupling ($\gamma\approx0$) and robust to instantaneous mixing ($\gamma\approx1$). Center–right: standard CT measures (center panel for CT1, i.e. $T_{xxxy}$, right panel for CT2, i.e. $T_{yxxx}$) are inflated by zero-lag mixing, the center panel shows a U-shaped dependence at intermediate values of $\gamma$, while the right panel displays a monotonic increase driven primarily by noise, as expected.}
  \label{fig:stat_overview}
\end{figure*}

We calculate for all CT1, CT2 and ACT measures the r-value (see Eq.\ref{eq:r}) for frequencies  between 1 Hz and 20 Hz and for all channel pairs. We used $N=100$ surrogate data sets.

\section{Results}
\subsection{Synthetic Data Experiment}
The experimental results obtained under the exploited generative model provide a strict confirmation of the expected qualitative behavior and illuminate the distinct responses of antisymmetrized and non-antisymmetrized metrics across the full mixing continuum parametrized by \(\gamma\). Below we summarize these findings in detail, referring to the left, center and right panels of the figure 3 (left: ACT, reported as median and IQR; centre and right: two non-antisymmetrized variants of CT, CT1 and CT2, respectively).

(i) \textbf{Genuine cubic coupling (\(\gamma \approx 0\)).} When the generative mechanism is dominated by true cubic interactions between the fundamental frequency \(f\) and its third harmonic \(3f\), both ACT and the standard (non-antisymmetrized) CT  $T_{xxxy}$ measures attain elevated values. This outcome is expected: genuine cubic coupling imposes structured, non-random phase relationships among the \(f\) and \(3f\) components that are not symmetric under channel permutation, hence the antisymmetrization step does not cancel the relevant signal. As a result, ACT preserves sensitivity to true higher-order harmonic dependencies and reports clear, statistically robust increases. The variant CT2, i.e. $T_{yxxx}$, depicts a qualitatively different behaviour. Indeed, CT2 was constructed deliberately as the counterpart required by our antisymmetrization scheme: it evaluates the presence of an interaction of the form that is not included in our generative model.

(ii) \textbf{Dominant instantaneous mixing / linear mixtures (\(\gamma \approx 1\)).} In the opposite regime, where instantaneous linear mixtures and shared noise sources dominate, the two non-antisymmetrized CT variants are systematically inflated. These measures pick up zero-lag correlations introduced by volume conduction and common noise, producing large apparent CT values even in the absence of any genuine non-linear, cross-frequency interaction. In contrast, antisymmetrization effectively cancels symmetric (zero-lag) contributions: ACT values remain close to baseline and negligible in this regime, demonstrating robust immunity to spurious inflation caused by instantaneous mixing.

(iii) \textbf{Intermediate mixing (\(0<\gamma<1\)).} For intermediate values of \(\gamma\) the non-antisymmetrized CT measures display a complex, confounded behavior. The centre panel shows a characteristic U-shaped dependence on \(\gamma\) for CT1 ($T_{xxxy}$; center panel of Figure 3), which reflects the competing influences of residual cubic structure and increasing instantaneous correlation: at low-to-moderate \(\gamma\) genuine cubic structure still elevates CT, while at higher \(\gamma\) shared noise again inflates CT, producing the observed U profile. Instead, CT2 ($T_{yxxx}$; right panel of Figure 3) depicts a qualitatively different monotonic increase for one specific CT variant, in which any increase observed in must arise from noise and instantaneous mixing rather than from a genuine cubic coupling.

\subsection{Real EEG Data Experiment}
\begin{figure*}[!t]      
  \centering
  \includegraphics[width=480pt]{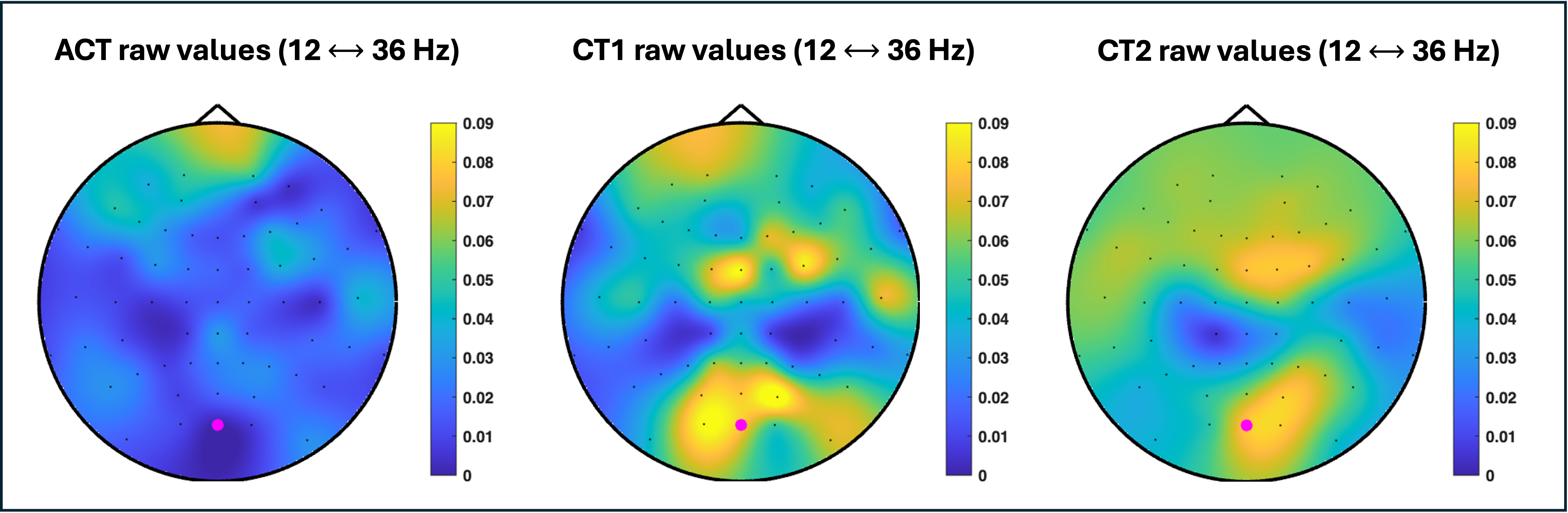} 
  \caption{Illustrative example of an empirical sensor-level seed map at 12 $\leftrightarrow$ 36 Hz using an occipital seed (magenta), shown for visualization of the spatial patterns discussed in the text. The ACT map (left panel) displays modest, spatially distributed increases toward frontal channels without short-range ``blooms''. In contrast, the CT1 and CT2 maps (center panel for CT1, i.e. $T_{xxxy}$; right panel for CT2, i.e. $T_{yxxx}$) exhibit pronounced, spatially confined high-value patches around the seed and its neighboring sensors, a pattern consistent with volume conduction rather than genuine long-range interactions.}
  \label{fig:stat_overview}
\end{figure*}
\begin{figure*}[!t]      
  \centering
  \includegraphics[width=480pt]{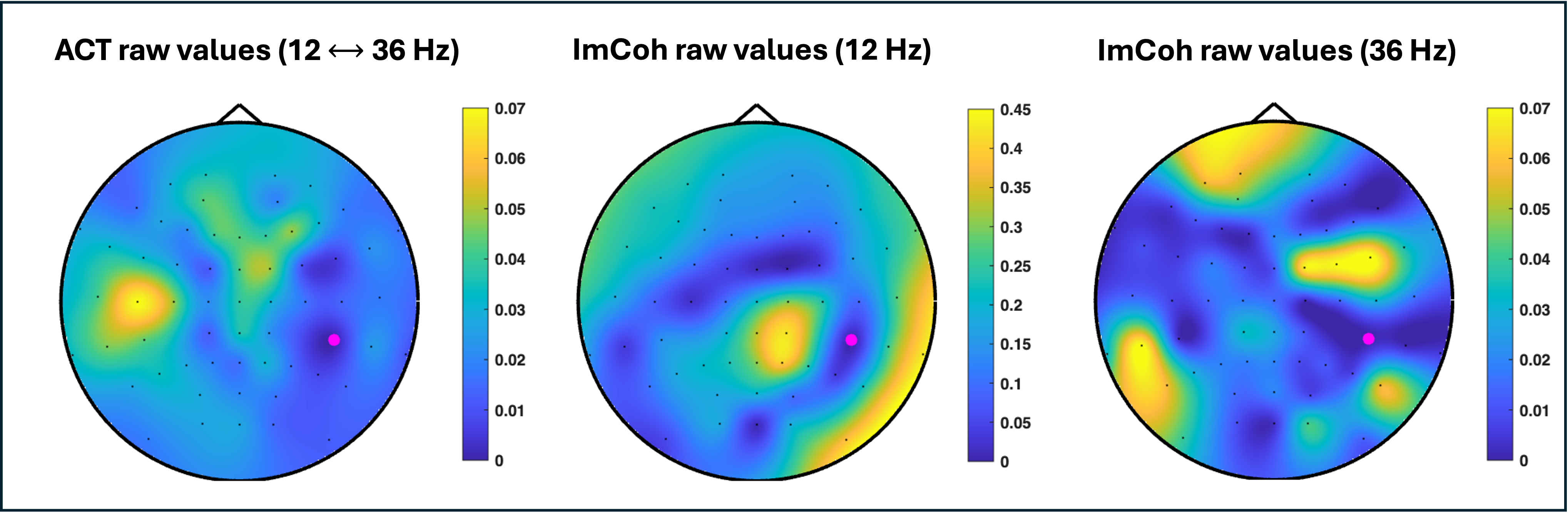} 
\caption{Illustrative example of an empirical sensor-level seed-based map at 12 $\leftrightarrow$ 36 Hz using a parietal seed (magenta), shown to visualize the long-range connectivity patterns discussed in the text. The ACT map (left panel) reveals pronounced inter-hemispheric connectivity at the single-subject level, a topology that is not reproduced by the corresponding ImCoh maps (center panel, 12 Hz; right panel, 36 Hz). The ACT pattern remains statistically robust after surrogate thresholding (Supplementary Figure S1), indicating that antisymmetric higher-order spectral metrics can reveal cross-frequency dependencies that conventional linear connectivity measures fail to capture.}
  \label{fig:stat_overview}
\end{figure*}
The results related to the real EEG data at single subject level align closely with the results obtained in simulations. As an example, we consider empirical sensor-level seed maps at 12$\leftrightarrow$36 Hz using an occipital seed. Both CT1 and CT2 display pronounced, spatially confined high-value patches centered on the magenta seed and its immediately adjacent sensors, a characteristic topography that is highly consistent with volume conduction, rather than genuine long-range interaction (center and right panels of Figure 4). In contrast, the ACT map does not exhibit these short-range “blooms” (left panel of Figure 4). Instead, it reveals only modest and spatially distributed increases toward frontal channels, suggesting a markedly different spatial organization once symmetric mixing contributions are removed.

Although this frontal ACT pattern could in principle reflect a genuine long-range cubic interaction between posterior and anterior regions, its overall magnitude remains small and, after surrogate-based thresholding (see Figure S1 in the Supplementary Materials), many of these values do not reach robust statistical significance. By contrast, several of the local CT peaks survive the same statistical testing, illustrating how non-antisymmetrized estimators can yield statistically significant yet artifactual connectivity driven by instantaneous mixing. This discrepancy emphasizes that statistical testing alone may not be sufficient when the estimator itself is sensitive to volume conduction. Similar results are observed when analyzing statistical dependencies at other frequencies. These observations reinforce the practical importance of employing antisymmetric, mixing-robust measures to meaningfully interpret high-order cross-frequency interactions at the sensor level, where linear mixing effects are otherwise pervasive and difficult to disentangle from genuine nonlinear neural dependencies.

To assess whether linear connectivity metrics can account for the findings obtained with a metric such as ACT, we perform a direct comparison between the results derived from imaginary part of coherency (ImCoh, \cite{Nolte2004}, which corresponds to a particular case of our general antisymmetric cross-polycoherence, with $m=2$, as shown in the subsection E of Section II) and those obtained with ACT. The statistical relationship between the connectome derived from the ACT (at frequencies $f_1=f_2=f_3=12$ Hz, resulting in $f_4=36$ Hz) and ImCoh was assessed using Pearson’s correlation. For the ImCoh at 12 Hz, we observed a correlation of $\rho=0.0437$ ($p<0.01$), while for the ImCoh at 36 Hz, the correlation was $\rho=0.0608$ ($p<0.001$). Although these results reach statistical significance due to the high number of degrees of freedom (3721 channel pairs), the extremely low correlation coefficients indicate that these metrics capture largely independent features of neural connectivity. Specifically, the cross-tricoherence explains less than 0.4\% of the variance of the ImCoh connectomes, suggesting a very limited capacity for these measures to account for one another and highlighting the distinct information provided by higher-order spectral analysis compared to linear phase-synchrony measures.

This conclusion may also be also visually evident. Figure~5 shows an ACT seed-based map obtained with a parietal seed, revealing a clear long-range inter-hemispheric pattern involving the contralateral sensorimotor region. In contrast, the corresponding ImCoh maps do not reproduce this topography. In particular, the long-range ACT pattern shown in Figure 5 is statistically robust after surrogate thresholding (see Supplementary Figure S1), further emphasizing that antisymmetric, higher-order spectral metrics disclose cross-frequency dependencies that standard linear connectivity estimates miss.

\section{Discussions and Conclusion}

In this work we introduce a family of antisymmetric cross-polyspectral indices that enable the systematic investigation of higher-order cross-frequency interactions across multiple time series. These nonlinear interactions are defined by the constraint that the sum of the frequencies associated with \(N-1\) source regions coincides with the frequency observed in the \(N\)-th region, which effectively serves as the output site where the nonlinear combination manifests. Interactions of this form are physiologically plausible and, as a proof of principle, can be observed even at the single-subject, sensor level in the datasets presented here.

The proposed indices are particularly timely given recent advances in multi-locus transcranial magnetic stimulation (mTMS). Unlike conventional TMS systems, mTMS employs overlapping coil arrays to enable rapid, multi-site stimulations without mechanical movement \cite{Koponen2018,Nieminen2022,Sinisalo2024}. By simultaneously engaging several partially overlapping cortical loci at distinct rates, such protocols may in principle generate composite spectral drives in regions of overlap. In these scenarios, restricting monitoring and control to single-frequency connectivity is likely to miss emergent cross-frequency phenomena that actually mediate the network-level effects of stimulation. Explicit estimation of such dependencies is therefore essential to inform and guide personalized, multi-site, multi-frequency stimulation strategies in the future.

Methodologically, the family of indices we derive both generalizes and complements existing connectivity measures (e.g., \cite{Nolte2004,Chella2016}), by providing a principled, antisymmetric construction that is robust to instantaneous linear mixing (volume conduction/field spread) and that scales naturally with interaction order.

Furthermore, to the best of our knowledge, current cross-frequency approaches are fundamentally unequipped to handle the conditions of Eq. (1) in a way that is robust to linear mixing and that includes more than two signals, even when considering simple scalar time series. The challenge becomes even more insurmountable for existing methods when moving toward multivariate and multidimensional scenarios. Specifically, there is no established infrastructure to investigate complex nonlinear couplings among $N$ interacting regions (multivariate) where each source is intrinsically vectorial (multidimensional), such as in the case of $N$ brain parcels each characterized by $m$ scalar components \cite{Basti2020}. Building upon the proposed indices to bridge this gap constitutes a logical and feasible next step.

Finally, beyond the methodological development presented here, our findings open practical avenues for future research. A primary objective involves validating these indices across larger subject cohorts.

While this study serves as a rigorous proof of concept, demonstrating that nonlinear cross-frequency interactions can be robustly identified even at the single-subject level, it raises questions regarding the generalizability and stability of these findings. Specifically, it remains to be determined whether the observed coupling patterns represent (inter-subjects) principles of cortical organization or if they exhibit a highly individual characterization. 

Furthermore, testing their utility as real-time markers for adaptive mTMS protocols will represent a pivotal step in translating these theoretical metrics into actionable tools for personalized neuromodulation. The availability of such robust metrics will be key to transitioning higher-order cross-frequency analysis from a theoretical tool to a practical driver for personalized network neuromodulation. A crucial requirement in this context is the ability to obtain reliable estimates from short data segments, typically on the order of minutes. Achieving such temporal resolution is essential for capturing the rapid dynamics of neural coupling. To this end, a rigorous statistical characterization, evaluating the trade-off between data length, signal-to-noise ratio, and spectral estimation techniques, is necessary to determine the most robust analytical approach for tracking connectivity states in near real-time \cite{Basti2020}.

\section*{Declaration of competing interest}
The authors declare that the research was conducted in the absence of any commercial or financial relationships that could be construed as a potential conflict of interest.

\section*{Acknowledgements}
AB is supported by the European Research Council (ERC Synergy) under the European Union’s Horizon 2020 research and innovation programme (ConnectToBrain; Grant Agreement No. 810377). The content of this article reflects only the author’s view and the ERC Executive Agency is not responsible for the content.
 
\newpage 
\renewcommand{\thefigure}{S\arabic{figure}}
\setcounter{figure}{0}
\begin{figure}
\centering
\includegraphics[width=0.48\textwidth]{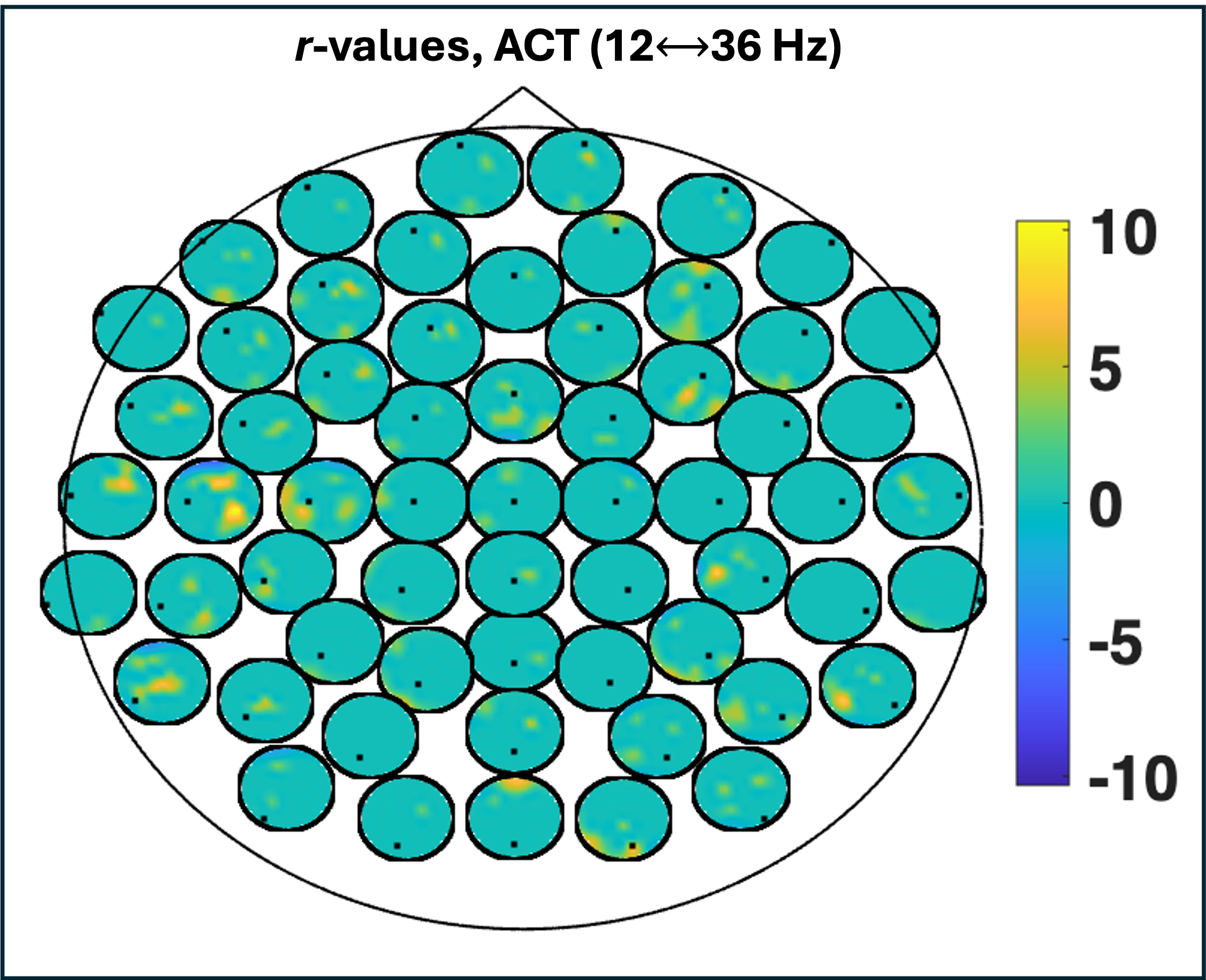}
\caption{Thresholded single-subject connectome for ACT, shown as a collection of seed-based maps displaying only statistically significant \(r\)-values.}
\end{figure}

\end{document}